\documentclass[aip,jcp,reprint,superscriptaddress,longbibliography]{revtex4-2}

\usepackage{amsmath,amssymb,bm}
\usepackage{graphicx}
\usepackage{physics}
\usepackage{hyperref}
\usepackage{color}
\usepackage{booktabs}
\usepackage{mathtools}
\usepackage{braket}

\usepackage{caption}
\usepackage{subcaption}

\usepackage{tikz}
\usepackage{pgfplots}
\usepackage{pgfplotstable}
\usetikzlibrary{arrows.meta,calc,positioning,patterns,decorations.pathreplacing}
\usepgfplotslibrary{groupplots}
\pgfplotsset{compat=1.17}
\usetikzlibrary{arrows.meta,calc,positioning}

\newcommand{\ii}{\mathrm{i}}

\begin{document}

\title{
Graph Symmetry Organizes Exceptional Dynamics in Open Quantum Systems
}

\author{Eric R. Bittner}
\affiliation{Department of Physics, University of Houston, Houston, Texas 77204, USA}

\author{Bhavay Tyagi}
\affiliation{Department of Physics, University of Houston, Houston, Texas 77204, USA}

\author{Kevin E. Bassler}
\affiliation{Department of Physics, University of Houston, Houston, Texas 77204, USA}

\date{\today}

\begin{abstract}
Exceptional points (EPs), indicative of parity–time (PT) symmetry breaking, play a central role in non-Hermitian physics, yet most studies begin from deliberately engineered effective Hamiltonians whose parameters are tuned to exhibit exceptional behavior.
In realistic open quantum systems, however, dynamics are governed by Lindblad superoperators whose spectral structure is high-dimensional, symmetry-constrained, and not obviously reducible to minimal non-Hermitian models.
A general framework for discovering exceptional dynamics directly from microscopic dissipative models has been lacking.
Here we introduce a symmetry-resolved approach for identifying and characterizing exceptional points directly from the full Liouvillian generator.
Correlated dissipation induces graph symmetries that decompose Liouville space into low-dimensional invariant sectors, within which minimal non-Hermitian blocks govern the onset of EPs and PT-breaking behavior.
We further introduce a numerical diagnostic — the exceptional-point strength $\mathcal{E}$ — based on eigenvector conditioning, which quantifies proximity to defective dynamics without requiring analytic reduction.
Applied to tight-binding models with correlated dephasing and relaxation, the method reproduces analytically predicted exceptional seams and reveals universal scaling of $\mathcal{E}$ near EP manifolds.
More broadly, the framework enables systematic discovery of hidden exceptional structure in complex or high-dimensional open systems and is naturally compatible with matrix-free and tensor-network implementations for scalable many-body applications.
\end{abstract}
\maketitle


\section{Introduction}
\label{sec:intro}

Exceptional points (EPs) have emerged as central organizing structures in non-Hermitian physics.
They generate enhanced sensitivity, non-analytic parameter dependence, and qualitative changes in dynamical behavior, and they underlie parity--time (PT) symmetry breaking across optics, condensed matter, and wave systems~\cite{Wiersig2020,Wiersig2014NatComm,ElGanainy2018}.
Foundational studies of non-Hermitian Hamiltonians~\cite{BenderBoettcher1998,Heiss2012}, topological classifications~\cite{BergholtzBudichKunst2021,AshidaGongUeda2020,Kawabata2019ClassificationEP}, and experimental realizations in photonics and sensing~\cite{MiriAlu2019,Naghiloo2019EP} have revealed the broad structural and technological importance of EPs.

In much of this literature, however, the starting point is an effective non-Hermitian Hamiltonian whose parameters are deliberately engineered or tuned to realize exceptional degeneracies.
Open quantum systems present a more fundamental setting: their dynamics are governed not by Hamiltonians alone, but by Liouvillian superoperators that encode coherent evolution, dissipation, and quantum jumps within a unified generator.
Liouvillian exceptional points therefore arise intrinsically within the full dynamical description rather than as reduced effective models~\cite{Minganti2019,Abo2024LEP,Seshadri2024JCP}.
Early work in chemical physics emphasized Liouvillian degeneracies in driven dissipative systems and highlighted limitations of approximate non-Hermitian reductions derived from nonequilibrium Green's functions~\cite{Mukamel2023JCP}, while subsequent studies clarified the role of dissipation channels in shaping Liouvillian EP structure~\cite{Arkhipov2020,Minganti2019}.

From a structural perspective, a Liouvillian $\mathcal{L}$ acting on density operators defines a linear dynamical map on a Hilbert--Schmidt operator space.
For finite-dimensional systems, $\mathcal{L}$ may be represented as a generally non-Hermitian $d^2 \times d^2$ matrix.
Trace preservation guarantees the existence of a left eigenvector corresponding to the identity operator and ensures at least one zero eigenvalue associated with stationary states.
However, these constraints do not prevent the generator from becoming defective.
A trace-preserving Liouvillian may exhibit nontrivial Jordan blocks at nonzero or even zero eigenvalues, producing algebraic time dependence and enhanced transient sensitivity despite conserving probability.

It is useful to distinguish rank deficiency from spectral defectiveness.
Rank deficiency reflects the existence of a nontrivial kernel and is typically associated with conserved quantities or steady states.
Defectiveness, by contrast, arises when the geometric multiplicity of an eigenvalue is smaller than its algebraic multiplicity, implying the presence of nontrivial Jordan blocks.
Exceptional points correspond to parameter values where eigenvalues and eigenvectors coalesce and defectiveness emerges.
In Liouville space, rank deficiency may signal protected or dark subspaces, whereas defectiveness governs the qualitative structure of transient dynamics.
Importantly, a Liouvillian may remain trace preserving and possess a unique stationary state while simultaneously exhibiting defective dynamics within nonzero spectral sectors.

The present work adopts a structural viewpoint on Liouvillian exceptional behavior.
Rather than assuming an effective non-Hermitian description \emph{a priori}, we develop a constructive diagnostic framework that identifies exceptional structure directly from the full Lindblad generator.
This perspective reveals that exceptional points need not be global properties of the Liouvillian spectrum.
Instead, when dissipation is correlated across system degrees of freedom, the generator admits a natural symmetry-resolved decomposition.

When correlated dissipation is represented as a graph acting on system degrees of freedom, the noise-correlation matrix defines an effective adjacency or Laplacian operator.
Its eigenvectors furnish irreducible representations of the dissipation network, and in this basis the dissipator decomposes into symmetry sectors.
Vanishing or near-vanishing dissipation eigenvalues correspond to modes that are weakly coupled to the environment.
Nontrivial dynamics arise when coherent or dissipative processes mix these symmetry sectors, generating minimal non-Hermitian blocks capable of supporting exceptional points or marginal oscillatory modes.

Within such symmetry-resolved representations, the full Liouvillian generically decomposes into low-dimensional invariant blocks.
Exceptional points therefore emerge within specific invariant subspaces rather than as global spectral singularities.
Sector localization has direct physical consequences: observables supported predominantly within a defective sector may exhibit algebraic prefactors or enhanced transient amplification even though the global steady state remains unique and stable.
Conversely, observables supported in diagonalizable sectors remain insensitive to such non-Hermitian criticality.
Exceptional behavior in open quantum systems is therefore revealed selectively through symmetry-resolved dynamical probes.

While invariant subspace decompositions of Lindblad generators are well known in abstract form~\cite{BaumgartnerNarnhofer2008JPA,AlbertJiang2014PRA}, we provide a constructive symmetry reduction tailored specifically to adjacency-matrix correlations induced by graph-structured dissipation.
This demonstrates explicitly that Liouvillian defectiveness is a symmetry-selected phenomenon localized within irreducible sectors of the traceless operator space.

Beyond specific models, we formulate a general computational workflow for mapping arbitrary tight-binding open systems onto symmetry-reduced Liouvillian blocks.
This enables systematic detection of protected subspaces, exceptional points, and synchronization-capable modes, transforming the identification of non-Hermitian dynamical phases from an abstract spectral problem into a practical structural analysis applicable to complex dissipative networks.

\section{Liouvillian Structure, Rank, and Defectiveness}
\label{sec:structure}

We consider Markovian open quantum systems governed by a time-independent
Lindblad master equation
\begin{equation}
\dot{\rho}(t) = \mathcal{L}[\rho(t)]
= -i[H,\rho] + \sum_{\mu} 
\left( L_\mu \rho L_\mu^\dagger - \frac{1}{2}\{L_\mu^\dagger L_\mu,\rho\} \right),
\end{equation}
where $\mathcal{L}$ is the Liouvillian superoperator.  

For a finite-dimensional Hilbert space with $\dim \mathcal{H} = d$, the density
matrix $\rho$ is a linear operator $\rho : \mathcal{H} \rightarrow \mathcal{H}$.
Upon vectorization, $\rho \mapsto |\rho\rangle$, the dynamics become linear,
\begin{equation}
\frac{d}{dt}|\rho(t)\rangle = \mathcal{L} |\rho(t)\rangle ,
\end{equation}
with $\mathcal{L} : \mathcal{B}(\mathcal{H}) \rightarrow \mathcal{B}(\mathcal{H})$.
In this representation, $\mathcal{L}$ is generally a non-Hermitian
$d^2 \times d^2$ matrix acting on the $d^2$-dimensional Hilbert--Schmidt space.
Its eigenvalues $\lambda_\alpha = -\Gamma_\alpha + i\omega_\alpha$ encode decay
rates and coherent oscillation frequencies.

Physical states $\{P \in \mathcal{B}(\mathcal{H}) \mid \mathrm{Tr}\,P = 1\}$
form an affine $(d^2 - 1)$-dimensional hyperplane in operator space.
Trace preservation implies
\begin{equation}
\frac{d}{dt} \mathrm{Tr}\,\rho(t) = \mathrm{Tr}(\mathcal{L}\rho) = 0 ,
\end{equation}
which is equivalent to the identity operator being a left zero-eigenvector of
the Liouvillian,
\begin{equation}
\langle I | \mathcal{L} = 0 , \qquad 
\langle I | \rho \rangle \equiv \mathrm{Tr}\,\rho .
\end{equation}
This condition constrains only the \emph{left} kernel of the generator.  It
places no restriction on the structure of the right eigenspaces, their
geometric multiplicities, or the diagonalizability of $\mathcal{L}$. Complete
positivity restricts the admissible spectral region but does not enforce
normality or spectral simplicity.

A useful structural decomposition follows from separating operator space into
traceful and traceless components,
\begin{align}
\mathcal{B}(\mathcal{H}) &= \mathrm{span}\{I\} \oplus \mathcal{B}_0(\mathcal{H}), \\
\mathcal{B}_0(\mathcal{H}) &= \{ X \in \mathcal{B}(\mathcal{H}) \mid \mathrm{Tr}\,X = 0 \}.
\end{align}
For any trace-preserving Liouvillian, the traceless subspace
$\mathcal{B}_0(\mathcal{H})$ is invariant under $\mathcal{L}$.
In an adapted basis the Liouvillian admits the block form
\begin{equation}
\mathcal{L} =
\begin{pmatrix}
0 & 0 \\
\mathcal{L}_{21} & \mathcal{L}_0
\end{pmatrix},
\label{eq:6}
\end{equation}
where $\mathcal{L}_0$ acts entirely within the traceless sector.

The existence of invariant operator subspaces under Lindblad evolution is a
general structural feature of quantum dynamical semigroups and is closely tied
to symmetry and conservation laws. Rigorous structural analyses show that
symmetries induce decompositions of Liouville space into invariant sectors
whose geometry governs degeneracies, conserved quantities, and steady-state
multiplicity\cite{BaumgartnerNarnhofer2008JPA,AlbertJiang2014PRA}. In this
sense, the block form of Eq.~\ref{eq:6} reflects the symmetry-resolved geometry
of the operator algebra.

All nontrivial spectral properties, degeneracies, and Jordan structure are
therefore governed by $\mathcal{L}_0$ and are logically independent of trace
preservation. Anomalous relaxation behavior may thus arise in traceless
observables even though $\mathrm{Tr}\,\rho(t)$ remains strictly conserved.

For a finite-dimensional linear operator, the rank is defined as
$\mathrm{rank}(\mathcal{L}) = \dim(\mathrm{Im}\,\mathcal{L})$
and is related to the nullity by the rank–nullity theorem.
Rank deficiency indicates the existence of a nontrivial kernel and loss of
invertibility. In Liouville space this typically arises from trace
preservation and the existence of steady states or conserved quantities.
Spectral defectiveness, by contrast, refers to the failure of diagonalizability
due to the appearance of nontrivial Jordan blocks at eigenvalue degeneracies
(exceptional points). The two notions are not equivalent: a Liouvillian may be
rank deficient yet diagonalizable, or spectrally defective without additional
rank loss if Jordan blocks occur at nonzero eigenvalues. Only when defectiveness
occurs at the zero eigenvalue do rank deficiency and exceptional points coincide.

A Liouvillian $\mathcal{L}$ is said to be \textit{defective} if, for some
eigenvalue $\lambda$, the geometric multiplicity
$\dim \ker(\mathcal{L}-\lambda I)$ is strictly smaller than the algebraic
multiplicity of $\lambda$. Equivalently, $\mathcal{L}$ is not diagonalizable
and possesses at least one nontrivial Jordan block. In this case the
Liouville spectrum includes generalized eigenoperators
$\{X_0, X_1, \ldots\}$ forming a Jordan chain satisfying
\begin{equation}
(\mathcal{L}-\lambda I)X_0 = 0, \qquad
(\mathcal{L}-\lambda I)X_1 = X_0 .
\end{equation}
Time evolution then acquires polynomial prefactors,
\begin{equation}
e^{\mathcal{L}t} X_1 = e^{\lambda t}\left(X_1 + t X_0\right),
\end{equation}
providing a direct dynamical signature of non-diagonalizable Liouville structure.
Trace preservation constrains only the left eigenvector associated with the
zero eigenvalue and does not preclude Jordan chains at nonzero $\lambda$.

Defectiveness can be detected in a basis-independent manner using
rank–nullity properties of powers of the shifted Liouvillian. For a candidate
eigenvalue $\lambda$, define
\begin{equation}
\delta_1(\lambda) = \dim \ker(\mathcal{L}-\lambda I), \qquad
\delta_2(\lambda) = \dim \ker\!\left[(\mathcal{L}-\lambda I)^2\right].
\end{equation}
A necessary and sufficient condition for the existence of at least one
nontrivial Jordan block at $\lambda$ is
\begin{equation}
\delta_2(\lambda) > \delta_1(\lambda).
\end{equation}
This provides a robust numerical diagnostic for identifying defective
Liouvillian dynamics without explicitly constructing a Jordan decomposition.
After symmetry reduction, the same test can be applied within each invariant
subspace, allowing defectiveness to be localized to specific symmetry sectors.
\section{Correlated Dissipation on Structured Networks}
\label{sec:structured}

We consider a general open quantum system consisting of $N$ local degrees of freedom coupled to a structured dissipative environment whose spatial or network correlations are encoded by a real symmetric matrix $\mathbf{A}$.  The matrix $\mathbf{A}$ may be interpreted as the weighted adjacency matrix of an abstract graph, but no restriction is imposed on its topology beyond the requirement of complete positivity of the resulting dissipator.  The elements $A_{ij}$ quantify the degree of bath correlation between sites $i$ and $j$, and determine how local dissipation channels are coherently or incoherently mixed by the environment.

Throughout this section we adopt the minimal correlated-noise model
\begin{equation}
\Gamma = I + c\, \mathbf{A},
\label{eq:Gamma_general}
\end{equation}
where $I$ is the identity matrix and $c$ controls the strength and sign of correlations ($c>0$ for correlated noise, $c<0$ for anti-correlated noise).  Complete positivity of the Lindblad generator requires $\Gamma \succeq 0$, which constrains the admissible range of $c$ according to the extremal eigenvalues of $\mathbf{A}$,
\begin{equation}
1 + c\,\lambda_{\min}(\mathbf{A}) \ge 0, \qquad
1 + c\,\lambda_{\max}(\mathbf{A}) \ge 0 .
\end{equation}
This bound defines the physically allowed correlation domain for any chosen network structure.

A central consequence of Eq.~\eqref{eq:Gamma_general} is that the Liouvillian inherits the symmetry group of $\mathbf{A}$.  The dissipator and, when compatible, the Hamiltonian decompose into invariant subspaces associated with the irreducible representations of the automorphism group of the underlying graph.  As shown in Sec.~\ref{sec:structure}, this symmetry resolution dramatically reduces the effective dimensionality of the Liouvillian and localizes nontrivial spectral structure—including exceptional points (EPs)—within small symmetry-selected blocks.
Importantly, this construction is completely general: any system whose environmental correlations admit a structured adjacency matrix may be analyzed in the same way.  The numerical diagnostics developed below therefore apply directly to arbitrary Liouvillian superoperators once their symmetry-resolved blocks are identified.

While invariant-sector decompositions of Lindblad generators are well
established in general\cite{BaumgartnerNarnhofer2008JPA,AlbertJiang2014PRA},
the present work does not propose a new classification theorem.
Rather, we provide a constructive identification of symmetry sectors
generated specifically by graph-structured correlated dissipation and
demonstrate how spectral defectiveness (exceptional points) can be
localized within these sectors.

Importantly, the symmetry sectors considered here are not imposed
externally at the level of the system Hamiltonian alone, but arise
constructively from the structure of the correlated dissipator
through its adjacency (or Laplacian) matrix.  In generic settings,
the Hamiltonian and dissipator need not share identical symmetry
representations, and exceptional behavior emerges precisely from
their sector-mixing interplay.  The present framework therefore
identifies defectiveness induced by graph-structured noise
correlations even in situations where Hamiltonian symmetry analysis
alone would not predict invariant dynamical blocks.

\paragraph{Adjacency spectrum and protected modes in $k$-regular graphs.}

For any connected $k$-regular graph on $N$ vertices, the adjacency matrix $A$ is real and
symmetric with a distinguished maximal eigenvalue $\lambda_{\max}=k$ and corresponding
normalized eigenvector $\mathbf{u}=(1,1,\ldots,1)/\sqrt{N}$.
This uniform mode represents the fully symmetric collective dissipation channel.
All remaining eigenvalues lie in a bounded interval $|\lambda_\alpha| \lesssim O(\sqrt{k})$
and, for generic or random regular graphs, form a dense spectral band centered near zero.

When the noise correlation matrix is written as $\Gamma=\gamma_0(I+cA)$, the dissipation
rates in each symmetry sector become
\begin{equation}
\gamma_\alpha = \gamma_0(1 + c\,\lambda_\alpha).
\end{equation}
Complete positivity imposes the bound $|c| \le 1/k$.
At the anti-correlated boundary $c=-1/k$, the symmetric mode satisfies
$\gamma_{\rm sym}=0$ and becomes strictly protected, producing a long-lived collective
channel capable of supporting marginal or limit-cycle dynamics when coherently mixed.
In contrast, all remaining adjacency eigenmodes retain strictly positive decay rates,
since their eigenvalues satisfy $|\lambda_\alpha| < k$.

Therefore, for generic $k$-regular graphs, the fully symmetric sector is the \emph{only}
irreducible representation that can become truly protected under correlated dissipation.
The remaining modes remain damped even at the positivity boundary, although near-degenerate
eigenvalues can still facilitate exceptional-point formation when coherent couplings mix
multiple sectors.

This clarifies that long-lived collective dynamics in regular networks arise uniquely from
anti-correlated noise and global symmetry, rather than from generic adjacency degeneracies.

\paragraph{Cycle graphs as a worked example.}
While the framework applies to arbitrary graphs, it is useful to illustrate these ideas using highly symmetric cases where analytic reduction is possible.  In the following, we use the cycle graph $C_N$ as a representative example.  Its translational symmetry admits an explicit Fourier decomposition that cleanly exposes symmetry-protected and weakly damped sectors, allowing the emergence of exceptional points to be analyzed transparently.

For the cycle graph $C_N$, the adjacency matrix is
\begin{equation}
A_{ij} = \delta_{i,j+1} + \delta_{i,j-1},
\end{equation}
with periodic boundary conditions.  Because $C_N$ is a $d$-regular graph with $d=2$, its spectrum satisfies $\lambda(A) \in [-2,2]$, implying the positivity constraint $c \in [-1/2,\,1/2]$.  The eigenvectors are discrete Fourier modes
\begin{equation}
u_k(j) = \frac{1}{\sqrt{N}} e^{\ii k j}, 
\qquad k = \frac{2\pi m}{N}, \quad m=0,\ldots,N-1,
\end{equation}
with corresponding eigenvalues
\begin{equation}
\Lambda(k) = 1 + 2c \cos k .
\label{eq:cycle_eigs}
\end{equation}
Each wavevector $k$ labels a symmetry sector of the dissipator.

The Fourier decomposition lifts directly to Liouville space through collective jump operators
\begin{equation}
\tilde{L}_k = \frac{1}{\sqrt{N}} \sum_{j=1}^N e^{\ii k j} L_j ,
\end{equation}
so that the dissipator separates into independent symmetry channels,
\begin{equation}
\mathcal{D}[\rho] = \sum_k \Lambda(k)\, \mathcal{D}_{\tilde{L}_k}[\rho].
\end{equation}
Density-matrix components transforming in different symmetry sectors do not mix under the dissipator alone.  When the Hamiltonian shares the same symmetry, the full Liouvillian block-diagonalizes into a direct sum of low-dimensional invariant blocks.

Modes satisfying $\Lambda(k) \approx 0$ correspond to symmetry-protected or weakly damped channels.  However, as emphasized in Sec.~\ref{sec:structure}, protection alone does not generate nontrivial dynamics; exceptional behavior emerges only when coherent or dissipative couplings mix distinct symmetry sectors.

\subsection{Minimal symmetry blocks and EP localization}

For finite sized system, the symmetry decomposition reduces the full Liouvillian to a direct sum of low-dimensional invariant blocks.  In many cases, the dynamics near an exceptional point are captured by minimal $2\times2$ or $3\times3$ non-Hermitian blocks formed by coupling a protected or weakly damped sector to a nearby dissipative sector.
The location and topology of exceptional points are therefore governed by the interplay between:
\begin{enumerate}
\item the spectrum of the correlation matrix $\mathbf{A}$,
\item the symmetry structure of the Hamiltonian couplings,
\item and the strength of symmetry-breaking perturbations.
\end{enumerate}
In the following sections we analyze these mechanisms explicitly for the two-site  model as a representative example, and demonstrate how the same numerical diagnostics can be applied to arbitrary Liouvillian superoperators.

\subsection{Stationary states, marginal modes, and limit cycles}
\label{sec:limit_cycles}

Beyond locating exceptional points in parameter space, the symmetry-resolved
formulation provides a direct route for characterizing the long-time dynamical
attractors of the system.  For any trace-preserving Liouvillian, at least one
right zero-eigenvector corresponds to a stationary density matrix.  In the dimer
model this stationary state resides entirely within the totally symmetric
operator sector and, for both dephasing and relaxation, approaches a mixed state
with vanishing coherences when dissipation dominates.

Symmetry plays a crucial role in determining whether additional slow or marginal
modes coexist with the stationary state.  In the absence of symmetry mixing, each
irreducible sector evolves independently, and protected sectors associated with
weakly damped dissipation channels remain dynamically inert.  When coherent or
dissipative couplings mix symmetry sectors, reduced Liouvillian blocks acquire
non-Hermitian structure capable of supporting exceptional points and nearly
marginal eigenmodes.

In the vicinity of an exceptional point, the real part of a reduced-block
eigenvalue approaches zero while its imaginary part remains finite, producing
long-lived oscillatory transients.  As the correlation strength approaches its
physical bound, $|c| \rightarrow c_{\max}$, the dissipation eigenvalue of a
protected sector vanishes, and the corresponding Liouvillian mode becomes
strictly marginal.  Time-domain propagation confirms that in this regime the
dynamics converge onto a persistent oscillatory trajectory in operator space
rather than relaxing to a static fixed point.  This behavior constitutes a
limit cycle of the density matrix, stabilized by symmetry-induced suppression of
dissipation rather than by fine tuning of system parameters.

Importantly, the symmetry decomposition allows these marginal and oscillatory
modes to be identified \emph{a priori} by examining which irreducible sectors
acquire vanishing dissipation eigenvalues and how they are coupled by the
Hamiltonian.  Rather than searching blindly in the full Liouville spectrum, one
may restrict attention to a small number of symmetry-selected blocks,
dramatically simplifying the detection of marginal dynamics and oscillatory
attractors.  This symmetry-based perspective provides the conceptual bridge to
the higher-dimensional network models analyzed below, where multiple protected
sectors coexist and the organization of limit cycles becomes even richer.

\paragraph{Static versus dynamical steady states.}
It is useful to distinguish between two inequivalent notions of ``steady state''
in open quantum dynamics.

A \emph{static steady state} is a fixed point of the Liouvillian,
\begin{equation}
\mathcal{L}\rho_{\rm ss} = 0,
\qquad
\rho(t) \rightarrow \rho_{\rm ss},
\end{equation}
corresponding to a time-independent density matrix.  In generic dissipative
systems this state is mixed and incoherent, and all initial conditions relax
exponentially toward this fixed point.  This scenario is implicitly assumed in
textbook-level treatments of open-system dynamics and an important fixed point
is the thermal density matrix.

By contrast, a \emph{dynamical steady state} corresponds to a stable periodic
attractor in Liouville space,
\begin{equation}
\rho(t) \rightarrow \rho_{\rm LC}(t),
\qquad
\rho_{\rm LC}(t+T) = \rho_{\rm LC}(t),
\end{equation}
such that all initial conditions converge onto the same oscillatory orbit.
Observable expectation values exhibit persistent oscillations even at long
times.  This behavior arises when the Liouvillian spectrum contains, in addition
to the stationary zero mode, a conjugate pair of purely imaginary eigenvalues,
\begin{equation}
\lambda_\pm = \pm i\omega ,
\end{equation}
while all remaining eigenvalues satisfy $\mathrm{Re}\,\lambda < 0$.

In this case the asymptotic state takes the form
\begin{equation}
\rho(t) =
\rho_0
+ c\, e^{i\omega t} R_+
+ c^* e^{-i\omega t} R_- ,
\end{equation}
where $R_\pm$ are the corresponding eigenoperators.  The attractor is therefore
not a fixed point but a one-dimensional invariant manifold (or higher-dimensional
torus for multiple marginal modes).  While the time-averaged density matrix may
appear mixed and incoherent, the instantaneous state carries persistent phase
coherence.  In previous work on correlated-noise synchronization, this structure
underlies spontaneous phase locking and collective oscillations.

It is important to emphasize that generic PT-broken Liouvillians do \emph{not}
satisfy this stronger marginality condition.  In most dissipative systems,
including the minimal models studied above away from the protection limit,
complex eigenvalues take the form
\begin{equation}
\lambda_\pm = -\gamma \pm i\omega, \qquad \gamma > 0,
\end{equation}
so that oscillations decay exponentially and the system ultimately relaxes to a
static mixed fixed point.  PT symmetry breaking in this sense signals oscillatory
\emph{transient} dynamics rather than true synchronization or dynamical steady
states.

The limit-cycle behavior observed as $|c| \rightarrow c_{\max}$ represents a
singular regime in which symmetry-induced protection drives the real parts of
selected Liouvillian eigenvalues exactly to zero.  This provides a concrete
mechanism for engineering dynamical steady states in correlated dissipative
networks, and serves as a precursor to the higher-dimensional examples analyzed
below.  Notably, this mechanism is robust across quantum statistics: bosonic,
fermionic, and anyonic dimers coupled to correlated environments all exhibit the
same symmetry-protected emergence of marginal Liouvillian modes and spontaneous
phase synchronization
\cite{BittnerTyagi2025JCP,TyagiLiBittner2024JPCL,BittnerTyagi2025SciRep}.

We now illustrate the general symmetry-resolved framework using minimal models in which all Liouvillian blocks can be constructed analytically.  The central point is that different physical dissipation mechanisms act on different operator sectors in Liouville space and therefore generate distinct reduced multiplicity blocks and distinct exceptional-point conditions, even when they are governed by the same underlying correlation matrix and graph spectrum.

\begin{figure}[t]
\centering
\includegraphics[width=0.95\linewidth]{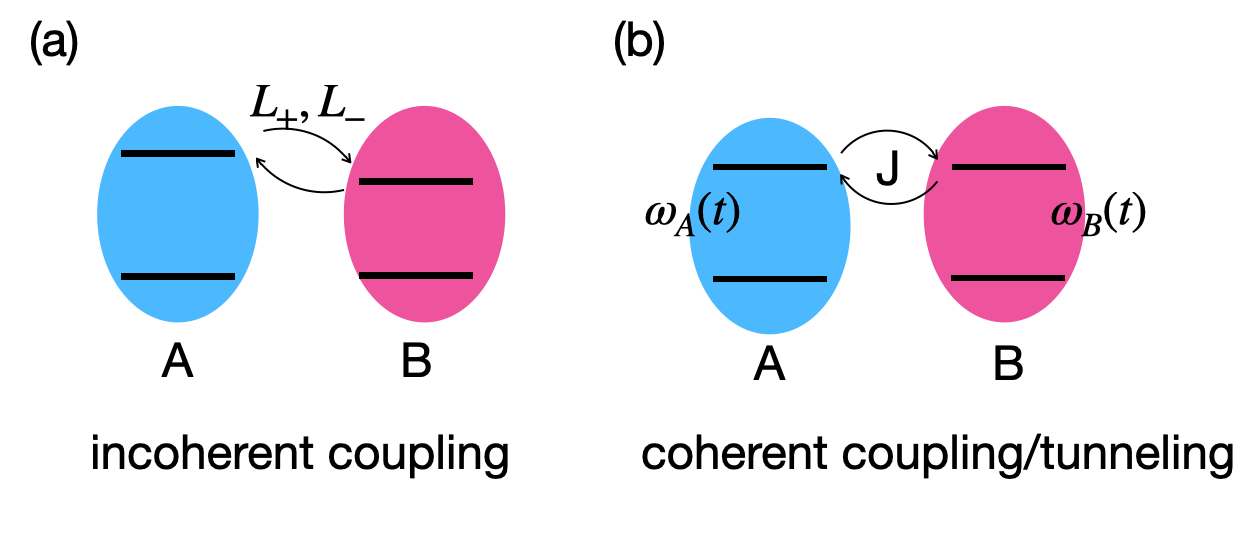}
\caption{
Two-site (dimer) models studied in this work.
\textbf{(a) Correlated relaxation (incoherent coupling).}
Each site $A$ and $B$ is a two-level system coupled to a common reservoir through symmetric and antisymmetric collective jump operators $L_\pm$, producing correlated decay channels that mix the single-excitation manifold. The sites may be detuned by an energy splitting $\Delta$ (not shown explicitly).
\textbf{(b) Correlated dephasing (coherent coupling / tunneling).}
The two sites are coherently coupled with tunneling amplitude $J$, while their local transition frequencies $\omega_A(t)$ and $\omega_B(t)$ undergo correlated stochastic fluctuations, generating collective dephasing without population transfer.
These minimal models isolate how coherent coupling and correlated dissipation reshape the symmetry-resolved Liouvillian spectrum, giving rise to exceptional points and non-Hermitian dynamical regimes.
}
\label{fig:dimer_models}
\end{figure}

\subsection{Distinct exceptional-point conditions for dephasing and relaxation
in a tunneling dimer}
\label{sec:examples}

Throughout this  section we focus on a two-site system as sketched in Fig.~\ref{fig:dimer_models}.
The two-site dimer may be viewed as a degenerate limit of a cycle graph, since its adjacency matrix coincides with the periodic tight-binding connectivity, although strictly speaking it is the path graph 
$P_2$; however, it captures the essential symmetry structure of larger rings.  The noise correlation matrix is taken to be
\begin{equation}
\Gamma = \gamma
\begin{pmatrix}
1 & c \\
c & 1
\end{pmatrix},
\label{eq:GammaDef}
\end{equation}
with eigenvalues $\gamma_\pm = \gamma(1 \pm c)$ corresponding to symmetric and antisymmetric collective channels with $c\in [-1,1]$. In the single-excitation subspace spanned by $\{|01\rangle,|10\rangle\}$,
we introduce the symmetric and antisymmetric states
\begin{equation}
|S\rangle = \frac{|01\rangle + |10\rangle}{\sqrt{2}}, \qquad
|A\rangle = \frac{|01\rangle - |10\rangle}{\sqrt{2}}.
\label{eq:SAbasis}
\end{equation}
These states furnish the irreducible representations of the exchange symmetry and provide a natural basis for symmetry reduction.

\vspace{0.5em}
Two qualitatively distinct classes of dissipative mechanisms play central roles:

\begin{enumerate}

\item \textbf{Correlated relaxation.}  
Local Lindblad operators take the form $L_j \propto \sigma^-_j$ or 
annihilation operators in the case of a bosonic system.  In this case the dissipator couples populations and coherences, and the effective Liouvillian blocks acquire additional non-Hermitian structure associated with irreversible excitation loss.  The symmetry selection rules governing which irreducible sectors become dynamically active differ from the dephasing case, leading to distinct exceptional-point conditions and scaling behavior.   This is sketched in Fig.~\ref{fig:dimer_models}(a) where the 
coupling to a correlated dissipative environment introduces incoherent hopping between sites A and B.

\item \textbf{Correlated dephasing (decoherence).}  
Local Lindblad operators take the form $L_J \propto \sigma^z_j$ (or number operators in bosonic realizations).  The dissipator suppresses coherences between symmetry sectors but does not directly exchange populations.  The resulting Liouvillian blocks governing coherence dynamics inherit the symmetry structure of the correlation matrix directly, and exceptional points arise through competition between coherent tunneling and symmetry-resolved dephasing rates.  This is sketched in Fig.~\ref{fig:dimer_models}(b) where the
coupling between sites A and B is introduced within the Hamiltonian via $J$ but the 
correlated environment acts to randomly modulate the local energy gaps, leading to dephasing.

\end{enumerate}

\begin{figure*}[t]
\centering

\begin{subfigure}[t]{0.48\textwidth}
\centering
\includegraphics[width=\linewidth]{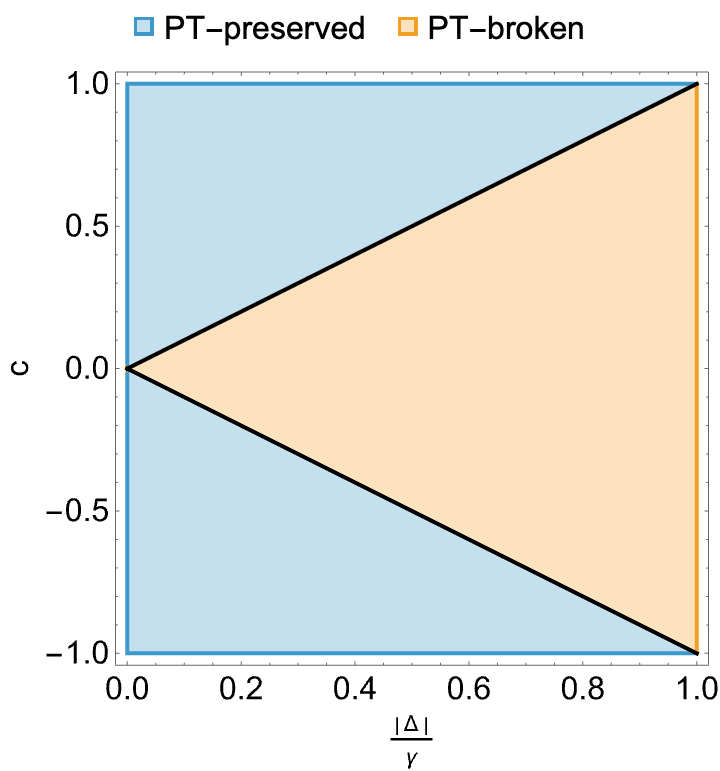}
\caption{Correlated relaxation in the $(|\Delta|/\gamma, c)$ plane.}
\label{fig:EP_relaxation}
\end{subfigure}
\hfill
\begin{subfigure}[t]{0.48\textwidth}
\centering
\includegraphics[width=\linewidth]{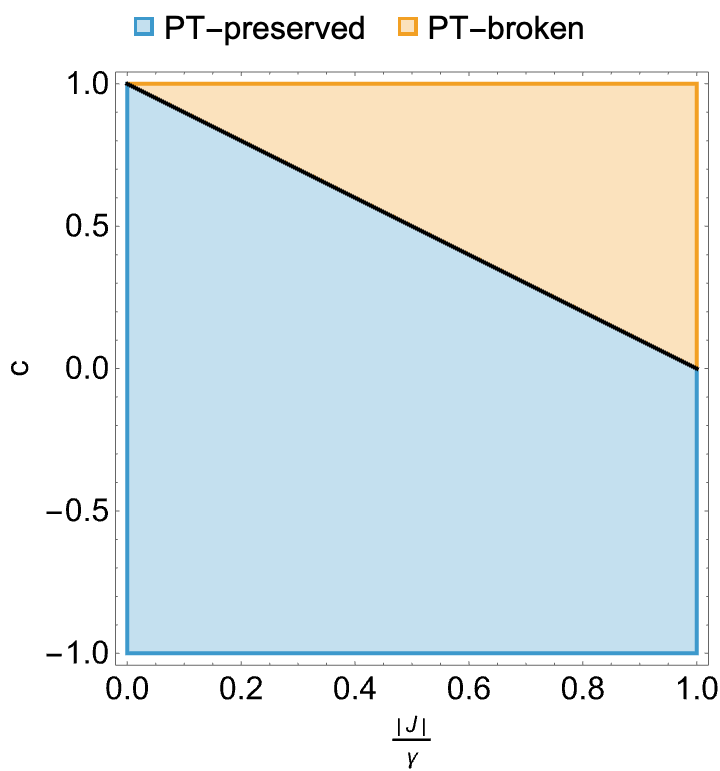}
\caption{Correlated dephasing in the $(|J|/\gamma, c)$ plane.}
\label{fig:EP_dephasing}
\end{subfigure}

\caption{
Exceptional-point phase diagrams for the two-site model with correlated noise.
Panels (a) and (b) show respectively the relaxation and dephasing cases.
Solid lines indicate the analytic exceptional-point boundaries derived from the reduced
generators $A_\Delta$ and $A_J$, respectively.
Shaded regions denote PT-preserved and PT-broken dynamical regimes as labeled.
}
\label{fig:EP_phase_diagrams}
\end{figure*}

Although both mechanisms originate from the same noise correlation matrix, the induced Liouvillian structures are not equivalent. This distinction is essential for identifying which symmetry sectors support exceptional-point (EP) behavior and how correlations reshape the dynamical phase landscape. 
Using the symmetric/antisymmetric basis defined in Eq.~\ref{eq:SAbasis}, the minimal nontrivial Liouville subspace is spanned by the real variables
\begin{equation}
\mathbf X = (y,z)^T, \qquad 
y = \Im \rho_{SA}, \qquad z = \rho_{SS} - \rho_{AA}.
\end{equation}
Projection of the master equation onto this sector yields a reduced non-Hermitian generator
\begin{equation}
\frac{d\mathbf X}{dt} = A_\alpha \mathbf X, \qquad \alpha \in \{J,\Delta\},
\end{equation}
whose spectral discriminant determines the exceptional-point condition.

\paragraph{Correlated relaxation.}

In the relaxation channel, the Lindblad operators take the form
$F_j = \sigma_j^-$ and the correlated bath induces symmetric and antisymmetric
collective decay modes,
\begin{equation}
L_\pm = \frac{1}{\sqrt{2}}\left(\sigma_1^- \pm \sigma_2^-\right),
\qquad
k_\pm = \gamma(1 \pm c),
\end{equation}
so that correlations directly modulate the imbalance between collective decay
rates. In contrast to the dephasing case, dissipation now couples both populations
and coherences, producing a qualitatively different non-Hermitian structure in
the symmetry-resolved dynamics.

Working  in the symmetric/antisymmetric basis of Eq.~\ref{eq:SAbasis}, the minimal
nontrivial Liouville sector is spanned by the real variables
\begin{equation}
\mathbf{X} = (y,z)^T, \qquad
y = \Im \rho_{SA}, \qquad z = \rho_{SS} - \rho_{AA},
\end{equation}
where $\rho_{SS}$ and $\rho_{AA}$ denote the populations of the symmetric and
antisymmetric single-excitation states. As shown in Appendix~\ref{app:A},
projection onto
this sector yields the reduced generator
\begin{equation}
\frac{d\mathbf{X}}{dt} = A_{\rm rel}\,\mathbf{X},
\qquad
A_{\rm rel} =
\begin{pmatrix}
0 & -\Delta \\
-\Delta & -\gamma(1-2c)
\end{pmatrix},
\label{eq:Arel}
\end{equation}
with correlation parameter $c \in [-1,1]$ and site-energy imbalance $\Delta$.

The reduced-block eigenvalues are
\begin{equation}
\lambda_\pm
= -\frac{\gamma(1-2c)}{2}
\pm \frac{1}{2}
\sqrt{\gamma^2(1-2c)^2 - 4\Delta^2},
\label{eq:lambda_rel}
\end{equation}
so that an exceptional point occurs when the spectral discriminant vanishes,
\begin{equation}
\gamma^2(1-2c)^2 - 4\Delta^2 = 0
\quad \Longleftrightarrow \quad
\frac{|\Delta|}{\gamma} = \frac{1}{2}\,|1-2c|.
\label{eq:EP_rel}
\end{equation}

For $|\Delta|/\gamma < \tfrac{1}{2}|1-2c|$, the eigenvalues are purely real and the
dynamics in this sector is PT-preserving, whereas for
$|\Delta|/\gamma > \tfrac{1}{2}|1-2c|$ they form a complex-conjugate pair and PT
symmetry is broken. In contrast to the dephasing channel, increasing correlations
enhance the effective imbalance between collective decay rates, promoting
non-Hermitian mixing between symmetry sectors and expanding the PT-broken region
of parameter space. Correlations therefore destabilize coherent symmetry sectors
under relaxation, yielding an EP topology that is inverted relative to the
dephasing case despite arising from the same underlying noise correlation matrix.

\paragraph{Correlated dephasing:}
As shown in Appendix~\ref{app:B}, projection of the master equation onto this sector yields the reduced generator
\begin{equation}
\frac{d\mathbf{X}}{dt} = A_{\rm deph}\,\mathbf{X},
\qquad
A_{\rm deph} =
\begin{pmatrix}
-\gamma(1-c) & -2J \\
+2J & 0
\end{pmatrix},
\label{eq:Adeph}
\end{equation}
with correlation parameter $c \in [-1,1]$.

The corresponding eigenvalues are
\begin{equation}
\lambda_\pm
= -\gamma(1-c)
\pm \sqrt{\gamma^2(1-c)^2 - 4J^2},
\label{eq:lambda_deph}
\end{equation}
so that an exceptional point occurs when the spectral discriminant vanishes,
\begin{equation}
\gamma^2(1-c)^2 - 4J^2 = 0
\quad \Longleftrightarrow \quad
\frac{|J|}{\gamma} = 1-c.
\label{eq:EP_deph}
\end{equation}
Equivalently,
\begin{equation}
c_{\rm crit} = 1 - \frac{|J|}{\gamma}.
\end{equation}

For $c < c_{\rm crit}$, the eigenvalues are purely real and the dynamics in this
symmetry-resolved sector is PT-preserving. For $c > c_{\rm crit}$, the eigenvalues
form a complex-conjugate pair and PT symmetry is broken. Increasing correlations
suppress the effective dephasing rate in the antisymmetric channel, thereby
stabilizing coherent oscillations and pushing the EP boundary toward larger
values of $|J|/\gamma$. In this sense, correlations act protectively in the
dephasing channel by enlarging the PT-preserved region of parameter space.

While the reduced generators $A_{\rm deph}$ and $A_{\rm rel}$ identify the local
eigenvalue bifurcations associated with exceptional points, a more complete
picture emerges by mapping the EP conditions over the full physically admissible
parameter space. Figure~\ref{fig:EP_phase_diagrams} shows representative phase
diagrams for correlated dephasing and correlated relaxation in the plane spanned
by the coherent control parameter and the correlation strength $c$.

For correlated relaxation [Fig.~\ref{fig:EP_phase_diagrams}(b)], the EP seam exhibits
a wedge-shaped topology governed by
\begin{equation}
\frac{|\Delta|}{\gamma} = \frac{1}{2}|1-2c|.
\end{equation}
Here correlations directly control the imbalance between collective decay
channels and therefore modulate the non-Hermitian mixing between populations and
coherences. At $c=0$, increasing detuning rapidly drives the system into the
PT-broken regime. By contrast, positive correlations enlarge the PT-preserved
region by reducing the effective decay imbalance, whereas negative correlations
enhance asymmetry between collective channels and promote exceptional behavior.
The relaxation phase diagram is therefore centered about $c=0$, in contrast to
the monotonic boundary observed in the dephasing channel.
Here, uncorrelated dissipation $(c=0)$ provides no symmetry protection against coherent detuning. Any finite energy imbalance $\Delta \neq 0$ immediately generates a complex-conjugate eigenvalue pair in the symmetry-resolved Liouvillian block, placing the system in the PT-broken phase. Finite bath correlations $(|c|>0)$ are required to stabilize a PT-preserved regime under detuning.

For correlated dephasing [Fig.~\ref{fig:EP_phase_diagrams}(a)], the EP seam follows
the linear condition
\begin{equation}
c_{\rm crit} = 1 - 2|J|/\gamma ,
\end{equation}
and separates a PT-preserved regime, in which the symmetry-resolved Liouvillian
spectrum remains purely real, from a PT-broken regime where a complex-conjugate
pair emerges. At weak coherent coupling ($|J|/\gamma \ll 1$), the system remains
PT-preserving over a broad range of correlations. Increasing tunneling competes
with symmetry-protected dephasing and progressively drives the system toward the
exceptional boundary. Positive correlations suppress effective dissipation in the
protected sector and stabilize coherent dynamics, while sufficiently strong
coherent mixing ultimately destabilizes this protection and induces PT symmetry
breaking.

Together, these phase diagrams demonstrate that identical noise-correlation
matrices generate inequivalent exceptional-point landscapes depending on the
physical dissipation mechanism. Correlations protect coherent symmetry sectors
under dephasing by suppressing effective dissipation, whereas under relaxation
they regulate population--coherence coupling through collective decay imbalance.
The topology and physical interpretation of the EP seams are therefore controlled
not by correlations alone, but by how correlations enter the symmetry-resolved
Liouvillian structure.

\paragraph*{Competing role of correlations.}
This analysis demonstrate explicitly that identical noise correlations generate inequivalent symmetry-resolved Liouvillian structures, depending on the underlying dissipation mechanism.In the dephasing channel, correlations suppress effective dissipation and stabilize coherent symmetry sectors. As a result, stronger coherent coupling, quantified by larger $|J|/\gamma$, is required to maintain PT symmetry.
In the relaxation channel, correlations enhance dissipative imbalance and destabilize coherent symmetry sectors; consequently, the critical detuning required to induce PT symmetry breaking shifts to smaller values of $|\Delta|/\gamma$, even though sufficiently strong detuning always drives the system into the PT-broken regime.



The preceding analysis establishes the central result of this section in  that correlated noise generates qualitatively distinct exceptional-point structures depending on whether dissipation acts through dephasing or relaxation channels. While these reduced models admit closed-form symmetry projections and analytic EP conditions, such constructions rapidly become impractical for larger Hilbert spaces, more complex connectivity graphs, or partially broken symmetries. In these settings, exceptional behavior must be identified directly from the Liouvillian spectrum and eigenmodes. We therefore turn to numerical diagnostics that quantify proximity to exceptional points through spectral clustering and eigenvector defectiveness, and illustrate their application in a higher-dimensional symmetry-resolved system.

\subsection{Exceptional-point strength and proximity scaling}
\label{sec:EP_strength}

Exceptional points (EPs) mark the breakdown of diagonalizability of a non-Hermitian generator and are accompanied by strong sensitivity of the dynamics to parameter variations.  
Rather than treating EPs solely as isolated bifurcation points, it is often more informative to quantify how close a system is to defective behavior throughout parameter space.  
To this end, we introduce a basis-independent measure of \emph{exceptional-point strength} that characterizes proximity to eigenvector coalescence directly at the level of the full Liouvillian.

Let $L$ denote a finite-dimensional non-Hermitian generator with right eigenvectors assembled into the matrix
\begin{equation}
V = \big( \mathbf{v}_1, \mathbf{v}_2, \ldots, \mathbf{v}_N \big),
\qquad
L V = V \Lambda ,
\end{equation}
where $\Lambda$ is the diagonal matrix of eigenvalues.  
At an EP the eigenvectors become linearly dependent and $V$ loses rank.  
Near an EP the eigenbasis becomes increasingly ill-conditioned, signaling the approach to defective dynamics.

We define the EP strength as
\begin{equation}
\boxed{
\mathcal{E}(L) \equiv \frac{1}{\sigma_{\min}(V)},
}
\label{eq:EPstrength}
\end{equation}
where $\sigma_{\min}(V)$ denotes the smallest singular value of the eigenvector matrix.  
Equivalently, $\mathcal{E}$ measures the inverse distance of $V$ from singularity in operator norm.  
This quantity diverges when $L$ becomes defective and remains finite for diagonalizable generators.  
The definition is invariant under similarity transformations of $L$ and does not depend on eigenvalue ordering or labeling.

For a generic second-order EP, perturbation theory predicts that the eigenvalue splitting and eigenvector coalescence scale as the square root of the distance to the EP in control-parameter space,
\begin{equation}
\lambda_{\pm}(\mu) \sim \lambda_{\mathrm{EP}} \pm \sqrt{\mu - \mu_{\mathrm{EP}}}, 
\end{equation}
implying
\begin{equation}
\sigma_{\min}(V) \sim |\mu - \mu_{\mathrm{EP}}|^{1/2},
\qquad
\boxed{
\mathcal{E}(\mu) \sim |\mu - \mu_{\mathrm{EP}}|^{-1/2}.
}
\label{eq:EPscaling}
\end{equation}
This universal scaling provides a quantitative signature of proximity to exceptional dynamics that can be extracted directly from the full Liouvillian spectrum.

\begin{figure*}[t]
\centering

\begin{subfigure}[t]{0.48\textwidth}
\centering
\includegraphics[width=\linewidth]{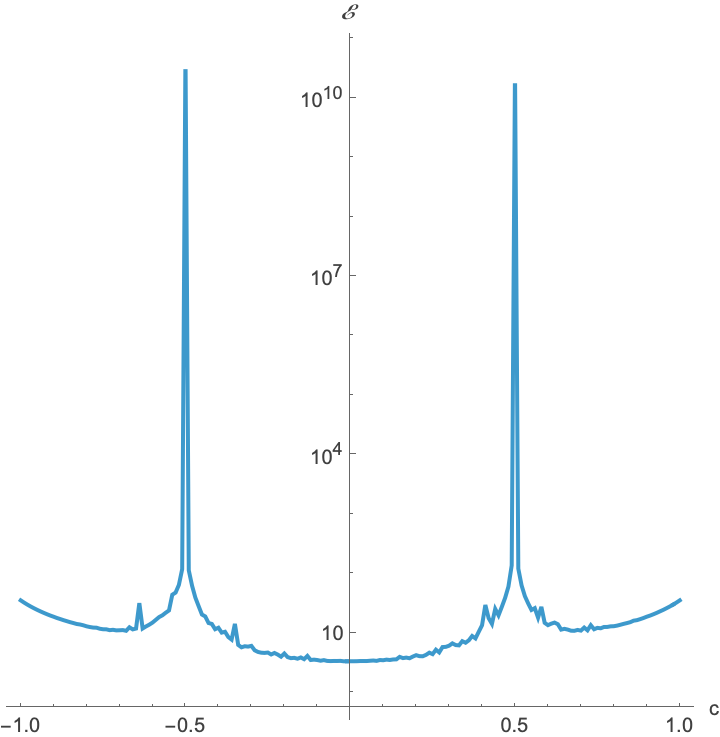}
\caption{Correlated relaxation: $\mathcal{E}$ versus $c$ at fixed $|\Delta|/\gamma$.
Two symmetric divergences appear, corresponding to the pair of exceptional points
predicted by the reduced-generator analysis.}
\label{fig:EP_strength_relax}
\end{subfigure}
\hfill
\begin{subfigure}[t]{0.48\textwidth}
\centering
\includegraphics[width=\linewidth]{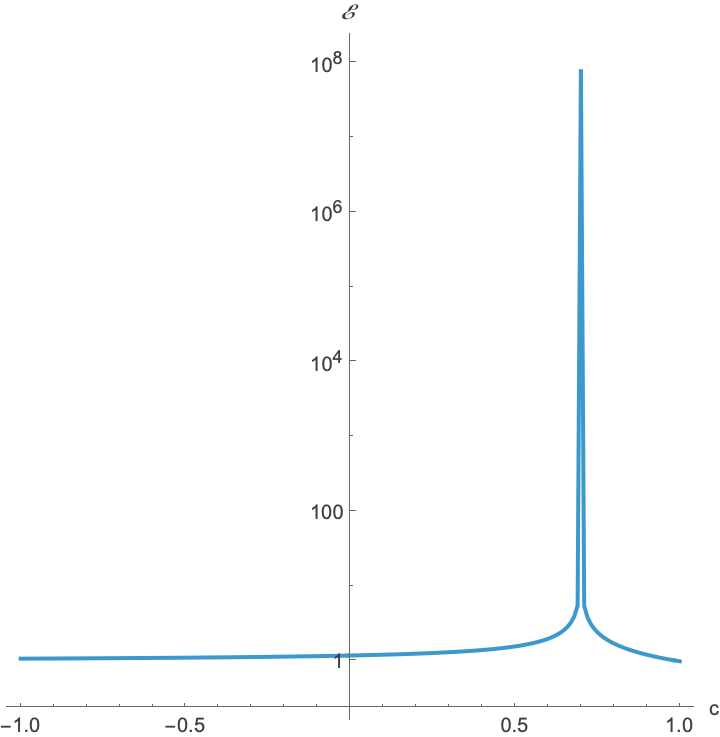}
\caption{Correlated dephasing: $\mathcal{E}$ versus $c$ at fixed $J/\gamma$.
A sharp divergence occurs at the analytically predicted exceptional-point condition
$c_{\mathrm{crit}} = 1 - |J|/\gamma$.}
\label{fig:EP_strength_deph}
\end{subfigure}

\caption{
Exceptional-point strength $\mathcal{E}$ [Eq.~\eqref{eq:EPstrength}] obtained from
one-dimensional parameter scans of the full Liouvillian.
In both panels, logarithmic scaling reveals the characteristic power-law growth
$\mathcal{E} \sim |\mu - \mu_{\mathrm{EP}}|^{-1/2}$ expected for second-order exceptional
points, confirming the square-root coalescence of eigenvectors near the EP seam.
Away from the seam, the EP strength remains finite, indicating a well-conditioned
eigenbasis.
These one-dimensional slices provide a quantitative measure of proximity to exceptional
dynamics and directly expose the sensitivity enhancement induced by nearby EP manifolds.
}
\label{fig:EP_strength_1D}
\end{figure*}

Figure~\ref{fig:EP_strength_1D} shows representative one-dimensional parameter scans of $\mathcal{E}$ for the correlated dephasing and correlated relaxation models.  
In each case, a single control parameter is varied while all other parameters are held fixed.  
The EP strength exhibits a sharp divergence precisely at the analytically predicted exceptional-point location.  
Away from the EP seam, $\mathcal{E}$ remains finite and slowly varying, indicating a well-conditioned eigenbasis and stable spectral structure.

On logarithmic axes, the growth of $\mathcal{E}$ near the peak is well described by the power-law scaling in Eq.~\eqref{eq:EPscaling}, consistent with the expected square-root coalescence of eigenvectors at a second-order EP.  
These results demonstrate that the EP strength provides a robust and sensitive diagnostic of proximity to defective Liouvillian dynamics, even when the system is not tuned exactly to the exceptional point.

For completeness, two-dimensional scans of $\mathcal{E}$ over parameter space are shown in Appendix~\ref{app:B}, illustrating how the EP strength recovers the full exceptional seam.  
While these maps validate the numerical detection procedure, the one-dimensional slices presented here provide a clearer quantitative view of the local scaling behavior and the sensitivity enhancement induced by nearby exceptional points.

More broadly, the EP strength furnishes a continuous measure of how strongly non-normal amplification and eigenvector coalescence influence the dynamics of open quantum systems.  
This opens the possibility of identifying latent exceptional structures in realistic models, where enhanced response, long-lived transients, or emergent oscillatory behavior may be governed by proximity to exceptional manifolds in parameter space rather than by fine-tuned EP crossings themselves.

\section{Discussion and physical outlook}
\label{sec:discussion}

The results developed here establish a unified framework for understanding how graph
symmetry organizes non-Hermitian dynamics in open quantum systems and for diagnosing the
emergence of exceptional behavior directly from the full Liouvillian generator.
Two complementary insights emerge.

First, symmetry decompositions induced by correlated dissipation naturally partition
Liouville space into low-dimensional invariant sectors in which minimal non-Hermitian
blocks control the onset of exceptional dynamics.
Within these blocks, correlated noise may either suppress or amplify dissipative
imbalance depending on the microscopic dissipation mechanism, producing qualitatively
distinct exceptional-point (EP) topologies for dephasing and relaxation channels.
This provides a transparent physical interpretation of how identical correlation
matrices can generate inequivalent dynamical phases once projected onto symmetry-adapted
subspaces.

Second, the exceptional-point strength $\mathcal{E}$ introduced in
Sec.~\ref{sec:EP_strength} furnishes a practical, model-independent diagnostic of
proximity to defective dynamics.
Because $\mathcal{E}$ is constructed solely from the eigenvector conditioning of the
full Liouvillian, it remains applicable even when analytic symmetry reduction is
impractical or when disorder, weak symmetry breaking, or additional couplings obscure
exact invariant subspaces.
The observed universal scaling of $\mathcal{E}$ near EP seams further enables quantitative
estimation of how strongly non-normal amplification influences system dynamics at finite
parameter detuning.

From a physical perspective, proximity to exceptional manifolds can have measurable
consequences even when the system is not tuned precisely to an EP.
Enhanced sensitivity, anomalously long-lived transients, amplified response to weak
perturbations, and slow decay of coherences are natural consequences of eigenvector
non-orthogonality in near-defective generators.
In this sense, EPs act as organizing centers in parameter space whose influence extends
well beyond the singular point itself.
The diagnostics developed here therefore provide a concrete route for identifying
``hidden'' or nearby exceptional structures that may shape experimentally observed
dynamics without being explicitly recognized as EP phenomena.

The methodology is broadly applicable to realistic open quantum platforms in which
correlated noise and network structure play an essential role.
Examples include coupled molecular excitonic aggregates and light-harvesting complexes,
where spatial correlations in environmental fluctuations influence transport and
collective coherence; solid-state spin networks and molecular qubits subject to spatially
correlated phonon baths; superconducting and photonic lattices with engineered dissipation;
and cavity-QED or polaritonic arrays where collective decay channels generate structured
Liouvillian spectra.
In such systems, microscopic modeling typically produces large Liouvillian generators
for which analytic spectral analysis is infeasible.
Automated EP detection via eigenvector conditioning provides a scalable computational tool
for surveying parameter landscapes, identifying protected or weakly damped modes, and
locating regimes of enhanced dynamical sensitivity.

A practical limitation of brute-force Liouvillian spectral searches is the rapid growth of
operator-space dimension: for a Hilbert space of dimension $d$, the Liouvillian acts on a
$d^2$-dimensional space and dense spectral routines scale prohibitively with $d^2$.
However, Lindblad generators arising from local Hamiltonians and structured dissipators
possess strong algebraic structure: $\mathcal{L}$ can be expressed as sparse sums of
Kronecker products and, in many-body settings, as matrix-product-operator (MPO) or
tensor-train (TT) superoperators.
This structure enables matrix-free Krylov methods and tensor-network eigensolvers to
target only the slow spectral sector relevant to exceptional dynamics, while symmetry
decomposition further restricts the search to low-dimensional invariant blocks.
Because the EP diagnostics rely only on repeated application of $\mathcal{L}$ rather than
dense diagonalization, they are naturally compatible with MPO/TT representations in which
both the density operator and Liouvillian are stored as low-rank tensor networks.
This provides a practical pathway for extending EP hunting to many-body open systems well
beyond the reach of explicit $d^2 \times d^2$ constructions.

More broadly, the framework developed here bridges three complementary perspectives:
graph-theoretic symmetry structure, non-Hermitian spectral topology, and open-system
dynamical stability.
By treating correlated dissipation as a symmetry-selecting graph and exceptional points as
localized instabilities within symmetry-adapted Liouvillian blocks, one obtains a
systematic route for diagnosing and engineering collective dynamical phases in complex
quantum networks.
Future directions include extension to higher-dimensional graphs, time-dependent driving
and feedback control, non-Markovian environments, and the exploration of nonlinear or
chaotic regimes in densely connected correlation networks.

Finally, it is important to emphasize that the present framework inverts the conventional
logic by which exceptional-point physics is typically explored.
Most studies of non-Hermitian dynamics begin with a deliberately constructed effective
Hamiltonian $H_{\mathrm{eff}}$ whose parameters are tuned to realize exceptional points or
PT transitions, and subsequently analyze spectral topology and dynamical consequences
within that assumed structure.
In contrast, our approach makes no \emph{a priori} assumption that a given open quantum
system admits a low-dimensional non-Hermitian reduction or even possesses exceptional
points at all.
Instead, the full Lindbladian is treated as the fundamental object, and exceptional
behavior is detected directly through numerical conditioning diagnostics in Liouville
space.

This perspective enables a systematic \emph{search} for exceptional dynamics in complex
models where analytic reduction is impractical or impossible, including systems with
disorder, competing dissipation channels, structured environments, or experimentally
constrained parameters.
Moreover, the diagnostics quantify proximity to exceptional manifolds, allowing one to
identify sensitivity enhancement and incipient PT-breaking behavior even when the system
is not tuned precisely onto an exceptional point.
In this sense, the method provides a general-purpose tool for uncovering hidden
non-Hermitian structure in realistic open quantum systems, rather than merely
characterizing models already known to exhibit exceptional points.

\begin{widetext}
\appendix
\section{ Dimer reduced generators and exceptional-point conditions}
\label{app:A}

\subsection{Variable mapping and symmetry-resolved coordinates}
\label{app:A_mapping}

In the single-excitation manifold spanned by $\{\ket{01},\ket{10}\}$ we use the
exchange-symmetry adapted basis
\begin{equation}
\ket{S}=\frac{\ket{01}+\ket{10}}{\sqrt{2}},
\qquad
\ket{A}=\frac{\ket{01}-\ket{10}}{\sqrt{2}}.
\end{equation}
In this $\{\ket{S},\ket{A}\}$ basis we define the real coordinates
\begin{equation}
p \equiv \rho_{SS}+\rho_{AA},
\qquad
z \equiv \rho_{SS}-\rho_{AA},
\qquad
y \equiv 2\,\Im(\rho_{SA}).
\label{eq:B_pzy_def}
\end{equation}
The pair $(y,z)$ forms a closed real subspace for both models considered below.
For pure dephasing, $p$ is conserved ($\dot p=0$), so the nontrivial dynamics
reduces to a $2\times2$ generator acting on $(y,z)^T$.

\subsection*{A.2. Exceptional points for correlated relaxation from the adjoint Liouvillian}
\label{app:A2_corr_rel_adj}

In this appendix we derive the exceptional-point (EP) condition for the correlated
relaxation model directly from the adjoint Liouvillian
$\mathcal{L}^\dagger$, in a manner consistent with the reduced generators
introduced in Eqs.~(28)--(32) of the main text. Because exceptional points are
defined by eigenvalue and eigenoperator coalescence of the Liouvillian
superoperator, the analysis must be carried out in operator space rather than
via affine state-space flow equations.

\paragraph{Model and collective decay channels.}
We consider the two-site model with correlated relaxation described by the jump
operators
\begin{equation}
L_\pm = \frac{1}{\sqrt{2}}(\sigma_1^- \pm \sigma_2^-),
\qquad
k_\pm = \gamma(1 \pm c),
\label{eq:A2_Lpm}
\end{equation}
with correlation parameter $c\in[-1,1]$, together with a site-energy imbalance
$\Delta$ that mixes the symmetric and antisymmetric single-excitation manifolds.
The adjoint Liouvillian acting on operators $O$ is
\begin{equation}
\mathcal{L}^\dagger[O]
=
i[H,O]
+
\sum_{\nu=\pm} k_\nu
\left(
L_\nu^\dagger O L_\nu
-
\frac{1}{2}\{L_\nu^\dagger L_\nu,O\}
\right).
\label{eq:A2_Ldagger}
\end{equation}

\paragraph{Symmetry-adapted operator basis.}
We work in the symmetric/antisymmetric single-excitation basis
\begin{equation}
|S\rangle = \frac{|10\rangle+|01\rangle}{\sqrt{2}},
\qquad
|A\rangle = \frac{|10\rangle-|01\rangle}{\sqrt{2}},
\label{eq:A2_SA_states}
\end{equation}
and introduce the Hermitian operators
\begin{equation}
Y = \mathrm{Im}(|S\rangle\langle A|),
\qquad
Z = |S\rangle\langle S| - |A\rangle\langle A|.
\label{eq:A2_YZ}
\end{equation}
This pair spans the minimal nontrivial symmetry-resolved operator sector relevant
for the correlated relaxation EP. Importantly, this sector is closed under
$\mathcal{L}^\dagger$ for the present model.

\paragraph{Adjoint-Liouvillian action.}
The Hamiltonian contribution produces coherent mixing between $Y$ and $Z$,
\begin{equation}
i[H,Y] = -\Delta\, Z,
\qquad
i[H,Z] = -\Delta\, Y,
\label{eq:A2_H_action}
\end{equation}
while correlated relaxation induces asymmetric damping between the symmetric and
antisymmetric sectors. Evaluating the dissipator terms in
Eq.~(\ref{eq:A2_Ldagger}) yields
\begin{equation}
\mathcal{L}^\dagger[Y] = -\Delta\, Z,
\qquad
\mathcal{L}^\dagger[Z] = -\Delta\, Y - \gamma(1-2c)\, Z.
\label{eq:A2_diss_action}
\end{equation}

\paragraph{Reduced adjoint generator.}
Combining Eqs.~(\ref{eq:A2_H_action}) and~(\ref{eq:A2_diss_action}), the evolution
of the operator vector $X=(Y,Z)^T$ is governed by
\begin{equation}
\frac{dX}{dt}
=
A_{\mathrm{rel}}\,X,
\qquad
A_{\mathrm{rel}}
=
\begin{pmatrix}
0 & -\Delta \\
-\Delta & -\gamma(1-2c)
\end{pmatrix},
\label{eq:A2_Arel}
\end{equation}
which is identical to Eq.~(30) in the main text.

\paragraph{Eigenvalues and exceptional-point condition.}
The eigenvalues of $A_{\mathrm{rel}}$ are
\begin{equation}
\lambda_\pm
=
-\frac{\gamma(1-2c)}{2}
\pm
\frac{1}{2}
\sqrt{\gamma^2(1-2c)^2 - 4\Delta^2},
\label{eq:A2_lambda}
\end{equation}
in agreement with Eq.~(31). An exceptional point occurs when the spectral
discriminant vanishes,
\begin{equation}
\gamma^2(1-2c)^2 - 4\Delta^2 = 0
\quad\Longrightarrow\quad
\frac{|\Delta|}{\gamma} = \frac{1}{2}|1-2c|,
\label{eq:A2_EP_cond}
\end{equation}
which reproduces Eq.~(32) of the main text.

\paragraph{Remarks.}
This derivation makes explicit that the correlated-relaxation EP arises from
defectiveness of a symmetry-resolved homogeneous adjoint-Liouvillian block. The
EP is therefore controlled by the imbalance between collective decay rates,
$k_+ - k_- = 2\gamma c$, and is not captured by affine state-space reductions that
eliminate this asymmetry. In the exactly correlated limit $c=1$, one decay channel
vanishes, leading to non-primitive dynamics that are distinct from the EP
identified above.
\subsection{Correlated dephasing: tunneling in the $\{\ket{S},\ket{A}\}$ basis}
\label{app:A_deph}

For correlated dephasing we take collective dephasing channels that diagonalize
the same correlation matrix, with rates
\begin{equation}
k_{\pm}=\gamma(1\pm c),
\qquad
c\in[-1,1],
\end{equation}
and consider coherent tunneling within the $\{\ket{S},\ket{A}\}$ block generated by
\begin{equation}
H_{\rm deph}=J\left(\ket{S}\!\bra{A}+\ket{A}\!\bra{S}\right).
\label{eq:B_Hdeph}
\end{equation}
In the $(y,z)$ sector, $p$ is conserved ($\dot p=0$), and the reduced dynamics is
purely homogeneous:
\begin{equation}
\frac{d}{dt}\binom{y}{z}
=
A_{\rm deph}\binom{y}{z},
\qquad
A_{\rm deph}=
\begin{pmatrix}
-2\gamma(1-c) & -2J\\
+2J & 0
\end{pmatrix}.
\label{eq:B_Adeph}
\end{equation}

The corresponding eigenvalues are
\begin{equation}
\lambda_\pm^{(\rm deph)}
=
-\gamma(1-c)
\pm
\sqrt{\gamma^2(1-c)^2-4J^2}.
\label{eq:B_deph_eigs}
\end{equation}
Thus the exceptional point occurs when
\begin{equation}
\gamma^2(1-c)^2-4J^2=0
\qquad\Longleftrightarrow\qquad
\gamma(1-c)=|J|
\qquad\Longleftrightarrow\qquad
c_{\rm crit}=1-\frac{|J|}{\gamma}.
\label{eq:B_deph_EP}
\end{equation}

For $c<c_{\rm crit}$ the reduced-block eigenvalues remain real (PT-preserved),
whereas for $c>c_{\rm crit}$ they form a complex-conjugate pair (PT-broken),
organizing the dephasing phase diagram in the $(|J|/\gamma,c)$ plane.

\section{Two-dimensional exceptional-point scans}
\label{app:B}

In this appendix we present two-dimensional parameter scans of the exceptional-point
strength $\mathcal{E}$ for the correlated dephasing and correlated relaxation models.
These scans serve as a global validation of the numerical EP detector introduced in
Sec.~\ref{sec:EP_strength} and demonstrate that the method faithfully reconstructs the
full exceptional seams predicted analytically.

For each point in parameter space, the full Liouvillian superoperator is constructed,
diagonalized, and the EP strength
\begin{equation}
\mathcal{E}(L) = \frac{1}{\sigma_{\min}(V)}
\end{equation}
is evaluated from the smallest singular value of the eigenvector matrix $V$.
Large values of $\mathcal{E}$ indicate proximity to defective dynamics, while smooth,
finite values correspond to well-conditioned spectral structure.

\subsection*{Correlated relaxation}

Figure~\ref{fig:EPscan_2D}(a) displays the corresponding scan for the correlated
relaxation model in the $(|\Delta|/\gamma, c)$ plane.
In contrast to dephasing, two symmetric ridges appear, reflecting the pair of exceptional
points generated by the reduced relaxation generator.
The numerical detector accurately reproduces the predicted seam geometry and confirms
the qualitative difference between the two dissipation mechanisms despite their shared
correlation matrix.

\subsection*{Correlated dephasing}

Figure~\ref{fig:EPscan_2D}(b) shows the EP strength for the correlated dephasing model
in the $(J/\gamma, c)$ plane.
A sharply localized ridge of enhanced $\mathcal{E}$ traces the analytically predicted
exceptional seam
\begin{equation}
c_{\mathrm{crit}} = 1 - \frac{|J|}{\gamma},
\end{equation}
terminating at the complete-positivity boundary $|c|=1$ for the two-site model.
Away from this ridge, the EP strength remains small and slowly varying, confirming that
no additional defective regions are present.

\begin{figure}[t]
\centering

\begin{subfigure}[t]{0.48\linewidth}
\centering
\includegraphics[width=\linewidth]{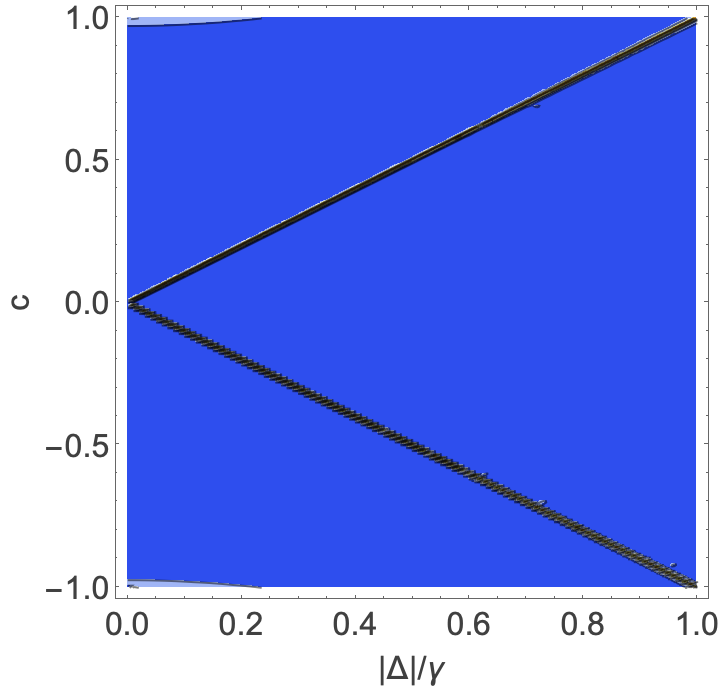}
\caption{Correlated relaxation in the $(|\Delta|/\gamma, c)$ plane. 
Two symmetric exceptional seams are resolved, consistent with the reduced relaxation generator.}
\label{fig:EPscan_relaxation_2D}
\end{subfigure}
\hfill
\begin{subfigure}[t]{0.48\linewidth}
\centering
\includegraphics[width=\linewidth]{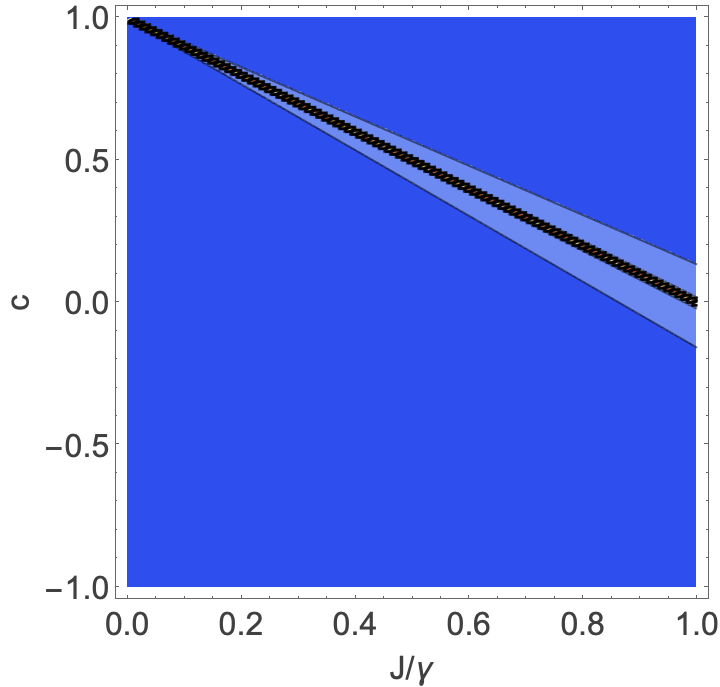}
\caption{Correlated dephasing in the $(J/\gamma, c)$ plane. 
A single sharply localized ridge traces the analytically predicted exceptional seam.}
\label{fig:EPscan_dephasing_2D}
\end{subfigure}

\caption{
Two-dimensional scans of the exceptional-point strength $\mathcal{E}$ for the dimer model.
Panels (a) and (b) correspond respectively to correlated relaxation and correlated dephasing,
illustrating the distinct topology of exceptional seams despite identical noise-correlation structure.
}
\label{fig:EPscan_2D}
\end{figure}

\subsection*{Remarks}

These two-dimensional maps demonstrate that the EP strength provides a reliable and
automated means of detecting exceptional manifolds directly from the full Liouvillian,
without reliance on symmetry reduction or analytical simplification.
However, because color-scale plots necessarily compress several orders of magnitude
into a limited dynamic range, the one-dimensional slices presented in
Sec.~\ref{sec:EP_strength} provide a more transparent view of the local scaling behavior
and the divergence structure near the exceptional seam.

Together, the global scans and the local scaling analysis establish the EP strength as a
practical diagnostic tool for identifying defective dynamical regimes in complex open
quantum systems.  
\end{widetext}

\section*{Supplementary Material.}
A fully documented \textit{Mathematica} notebook implementing the dimer models, Liouvillian construction, exceptional-point detection metrics, and numerical parameter scans used in this work is available online at
\url{https://www.wolframcloud.com/obj/bittner/Published/EP_detect_Dimer_example.nb}.
The notebook reproduces the numerical results reported in the main text and appendices.

\section*{Data Accessibility Statement}
All data generated by this work are provided in the manuscript or in the Supporting Information.

\begin{acknowledgments}
The work at the University of Houston was supported by the National Science Foundation under CHE-2404788 and the Robert A. Welch Foundation (E-1337).
\end{acknowledgments}

\section*{Author Contribution Statement}
ERB conceived the project, developed the theoretical framework,
and carried out the analytical and numerical calculations.
BT was involved in developing the theoretical framework
and editing the final manuscript.
KEB was involved with the conceptualization of the project, 
discussion of the results, 
and framing the idea in the general scope of graph theory. 
Generative AI tools were used during manuscript preparation to assist with drafting, structural organization, and verification of intermediate algebraic steps. All theoretical formulations, derivations, simulations, and physical interpretations were independently developed and validated by the authors, who take full responsibility for the work.

\bibliography{references}

\end{document}